\documentclass[english,10pt,aps,pra,twocolumn,superscriptaddress,floatfix]{revtex4-1}

\usepackage[T1]{fontenc}
\usepackage{amsmath}
\usepackage{amssymb}
\usepackage{amsfonts}
\usepackage{bm}
\usepackage{bbm}
\usepackage{epsfig}
\usepackage{grffile}
\usepackage{graphics}
\usepackage{amsthm}

\usepackage[usenames,dvipsnames]{color}
\definecolor{dblue}{rgb}{0,0,.5}

\usepackage[colorlinks=true,citecolor=dblue,linkcolor=dblue,urlcolor=dblue]{hyperref}
\usepackage[all]{hypcap}

\newcommand{\ud}{\mathrm{d}}

\newcommand{\hid}{\hat{\mathbbm{1}}}

\newcommand{\mc}[1]{\mathcal{#1}}

\renewcommand{\H}{\mc{H}}
\newcommand{\hH}{\hat{H}}
\newcommand{\hh}{\hat{h}}
\newcommand{\hs}{\hat{\sigma}}
\newcommand{\hU}{\hat{U}}
\newcommand{\hu}{\hat{u}}
\newcommand{\hO}{\hat{O}}
\newcommand{\hP}{\hat{P}}
\newcommand{\hQ}{\hat{Q}}
\newcommand{\hR}{\hat{R}}
\newcommand{\hV}{\hat{V}}
\newcommand{\veps}{\varepsilon}
\newcommand{\tveps}{{\tilde{\varepsilon}}}
\newcommand{\bomega}{{\bar{\omega}}}

\newcommand{\N}{\mc{N}}
\newcommand{\bN}{\bar{\mc{N}}}
\newcommand{\A}{\mc{A}}
\newcommand{\B}{\mc{B}}
\newcommand{\Ht}{\tilde{\mc{H}}}
\renewcommand{\NG}{{N_g}}
\newcommand{\crct}{\text{circ}}
\newcommand{\Q}{\mc{Q}}
\newcommand{\M}{\mc{M}}
\newcommand{\vol}{\operatorname{vol}}
\newcommand{\diag}{\operatorname{diag}}

\newtheorem{theorem}{Theorem}
\newtheorem{lemma}{Lemma}

\newcommand{\duke} {Department of Physics, Duke University, Durham, North Carolina 27708, USA}
\newcommand{\dukeM} {Department of Mathematics, Duke University, Durham, North Carolina 27708, USA}

\newcommand{\Title} {Fundamental limitations for measurements in quantum many-body systems}
\newcommand{\Authors}
{
\author{Thomas Barthel}
\affiliation{\duke}
\author{Jianfeng Lu}
\affiliation{\dukeM}
\affiliation{\duke}
}
\newcommand{\Date} {February 2, 2018}

\begin{document}

\title{\Title}
\Authors

\begin{abstract}
Dynamical measurement schemes are an important tool for the investigation of quantum many-body systems, especially in the age of quantum simulation. Here, we address the question whether generic measurements can be implemented efficiently if we have access to a certain set of experimentally realizable measurements and can extend it through time evolution. For the latter, two scenarios are considered (a) evolution according to unitary circuits and (b) evolution due to Hamiltonians that we can control in a time-dependent fashion. We find that the time needed to realize a certain measurement to a predefined accuracy scales in general exponentially with the system size -- posing a fundamental limitation. The argument is based, on the construction of $\veps$-packings for manifolds of observables with identical spectra and a comparison of their cardinalities to those of $\veps$-coverings for quantum circuits and unitary time-evolution operators. The former is related to the study of Grassmann manifolds.
\end{abstract}

\date{\Date}
\maketitle

\section{Introduction}\label{sec:intro}
In experiments with quantum many-body systems, we usually have direct access only to a relatively small set of standard observables in measurements. For quantum computation devices, these are often Pauli measurements.
In ion-trap systems, state-dependent laser-induced resonance fluorescence allows for the measurement of qubits in the computational basis \cite{Cirac1995-74,Leibfried2003-75,Myerson2008-100}.
For superconducting qubits, such projective measurements can be realized through a state-dependent shift in the resonance frequency of a dispersively coupled cavity \cite{Blais2004-69,Wallraff2004-431}. Also projection operators onto specific multi-qubit product states have been measured \cite{DiCarlo2010-467}.
For ultracold atoms, the particle density can be accessed through absorption imaging \cite{Bloch2007} and more recently developed quantum gas microscopes with single-site resolution based on fluorescence imaging \cite{Bakr2009-462,Sherson2010-467,Haller2015-11}.

Dynamical control can be used to measure observables that are not directly accessible. This is especially important for the purpose of quantum simulation \cite{Manin1980,Feynman1982-21,Lloyd1996-273,Kliesch2011-107}. The design of quantum simulators is advancing rapidly \cite{Castelvecchi2017-541,Gross2017-357,Bernien2017-551,Brown2016-2,Bohnet2016-352,Zhang2017-551}.
The relevant observables for the simulated systems will often not be directly accessible in the simulating device and hence require dynamical measurement schemes. While the investigation of general abilities and limitations of such schemes has just begun, several particular incarnations are successfully used in experiments:

Measurement of Pauli-$\hs_x$ and $\hs_y$ for ion-trap qubits are realized through the application of single-qubit gates and subsequent measurement of $\hs_z$. More elaborate schemes employ two-qubit gates, spin echo, spatial shuttling of qubits or hiding in non-computational electronic states, e.g., to do Bell-state measurements \cite{Riebe2004-429,Barrett2004-429}.
Similarly, for superconducting circuits, Bell-state measurements can be realized \cite{Steffen2013-500} through application of single-qubit rotations and controlled phase gates \cite{DiCarlo2009-460} before the standard Pauli measurements.
In ultracold atom experiments, the momentum distribution is obtained in time-of-flight measurements by letting the quantum gas expand freely before absorption imaging \cite{Anderson1995-269,Davis1995-75,Bloch2007}. Double-occupancies can be determined by rapid ramping of the lattice potential, tuning of interaction strengths, mapping double occupancy to a previously unpopulated spin state using radio-frequency pulses, and final absorption imaging \cite{Joerdens2008-455,Strohmaier2010-104}. Nearest-neighbor correlations have been measured through an additional modulation of the lattice depth or deformation of a superlattice
\cite{Trotzky2010-105,Greif2011-106,Greif2013-340}.
Bloch band populations can be examined by adiabatic band mapping \cite{Kastberg1995-74,Greiner2001-87,Koehl2005-94}.
Solid-state materials are studied with various scattering and microscopy techniques. The control over the Hamiltonian is naturally rather limited in this case. Nevertheless, pump-probe schemes are, for example, employed in time-resolved optical and photoemission spectroscopy \cite{Luo2016-3,Cavalieri2007-449,Smallwood2016-115}, scanning tunneling microscopy \cite{Cocker2013-7,Yoshida2014-9}, and electron microscopy \cite{Adhikari2017-9,Feist2017-176,Hassan2017-11} to enlarge the set of accessible observables.

In principle, arbitrary observables can be evaluated after state tomography \cite{Hradil1997-55,BlumeKohout2010-12} or compressed sensing procedures \cite{Gross2010-105}. However, for many-body systems, the number of required measurements and classical computational resources grow exponentially with increasing system size unless additional strong constraints can be leveraged \cite{Toth2010-105,Cramer2010-1}.

\begin{figure}[t]
  \includegraphics[width=\columnwidth]{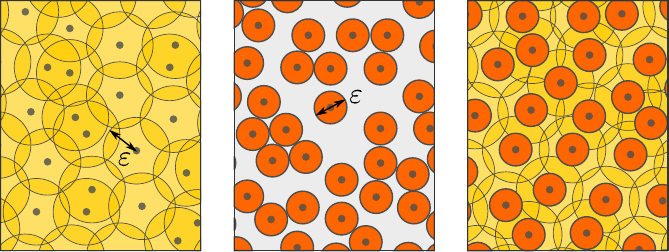}
  \caption{\label{fig:coversPacksNets}Left: An \emph{$\veps$-covering} $\Q$ for a space $\M$ with metric $d$ is a subset of $\M$ such that for every $z\in\M$ there is an $x\in\Q$ with $d(x,z)\leq\veps$. The cardinality $\N(\M,d,\veps)$ of the smallest $\veps$-covering is called the \emph{$\veps$-covering number}
  of $(\M,d)$.
  We are comparing covering numbers for sets of evolved observables $\hU^\dag\hO\hU$ with covering numbers for the set of all observables with the same spectrum as $\hO$.
  Center: An \emph{$\veps$-packing} $\Q$ is a subset of $\M$ such that all $x\neq y\in\Q$ have distance $d(x,y)>\veps$. Right: An \emph{$\veps$-net} is an $\veps$-covering and, at the same time, an $\veps$-packing.}
\end{figure}

Here, we assess the efficiency of dynamical measurement schemes by derivation of lower bounds on covering numbers for manifolds of observables with identical spectra (sections~\ref{sec:projections} and \ref{sec:observables}) and by comparing them to upper bounds on covering numbers for quantum circuits and unitary time-evolution operators (sections~\ref{sec:circuit} and \ref{sec:tevol}). Note that $\veps$-covering numbers $\N(\veps)$ and $\veps$-packing numbers $\bN(\veps)$ of a metric space are closely related with $\bN(2\veps)\leq \N(\veps)\leq\bN(\veps)$ \cite{Kolmogorov1959-14,Shiryayev1992} (Fig.~\ref{fig:coversPacksNets} and Appx.~\ref{sec:CoveringsPackings}). The analysis shows that the time needed to realize a certain measurement to a predefined accuracy scales in general exponentially with the system size. The result holds for the spectra of all typical observables of many-body systems.
In the following, we consider lattice systems consisting of $L$ $d$-dimensional qudits. We use the Bachmann-Landau symbols $\mc{O}$ and $\Omega$ for upper and lower bounds. Proofs for the lemmas are given in appendices. Variants of lemmas \ref{lemma:CoverUn},\ref{lemma:CoverProduct},\ref{lemma:CoverQuotient}, and \ref{lemma:CoverLipschitz} have been stated by Szarek in Refs.~\cite{Szarek1981,Szarek1983-151}.

\section{Evolution due to unitary quantum circuits}\label{sec:circuit}
First, let us consider the case of observables that are evolved using unitary circuits composed of arbitrary $k$-site gates.
\begin{theorem}\label{theorem:Circuit}
  Let $\hO=\hO^\dag$ be an observable and consider quantum circuits of size $\NG>L$ with each gate $\hu_i$ acting on at most $k$ sites. The evolved observables $\{\hU^\dag\hO\hU\}$ with any such quantum circuits $\{\hU=\prod_{i=1}^\NG\hu_i\}$ are elements of
  \begin{equation}\label{eq:Ncircuit}
  	\N_\crct\leq L^{k \NG}\left(\frac{14\NG}{\veps}\right)^{d^{2k}\NG}
  	=e^{\mc{O}(\NG\ln \NG)}
  \end{equation}
  balls of radius $\veps w(\hO)$ in operator space. Here, $w(\hO):=(\omega_{\max}-\omega_{\min})/2$ denotes the spectral width of $\hO$, i.e., half the difference of the maximum and minimum eigenvalues of $\hO$.
\end{theorem}
This can be shown by first bounding covering numbers for the quantum circuits $\hU$. In similar situations, Refs.~\cite{Poulin2011-106b,Kliesch2011-107} approximated the $k$-qudit gates $\hu_i$ by small circuits built from a finite gate library. This can be done as in practical implementations for quantum computation by first decomposing them into single-qubit and CNOT gates
\cite{Barenco1995-52,Moettoenen2004-93,Shende2006-25} and further approximating the latter according to the Solovay-Kitaev algorithm \cite{Kitaev2002,Dawson2006-6} or alternative schemes \cite{Kliuchnikov2013-110}.
However, one can take a more direct approach and simply employ an $\tveps$-covering for the $k$-qudit gates.
\begin{lemma}\label{lemma:CoverUn}
  For $0<\tveps\leq 1/10$, the $\tveps$-covering number for the unitary group $U(n)$ with respect to the operator-norm distance obeys
  \begin{equation}
  	\left(\frac{3}{4\tveps}\right)^{n^2} \leq \N\left(U(n),\|\cdot\|,\tveps\right) \leq \left(\frac{7}{\tveps}\right)^{n^2}.
  \end{equation}
\end{lemma}

We fix an $\tveps$-covering $\Q$ for the set $U(d^k)$ of all gates.
For a circuit $\hU=\prod_{i}\hu_i$, let $\hU_\tveps$ be the circuit where each of the $\NG$ gates is replaced by the nearest element in $\Q$. Then, according to the triangle inequality, $\|\hU_\tveps-\hU\|\leq \NG\tveps$ and, choosing $\tveps=\veps/(2\NG)$,
\begin{equation}
	\|\hU_\tveps^\dag\hO\hU_\tveps-\hU^\dag\hO\hU\|\leq 2\|\hU_\tveps-\hU\| w(\hO)\leq \veps w(\hO).
\end{equation}
The upper bound in lemma~\ref{lemma:CoverUn} gives $|\Q|\leq (14\NG/\veps)^{d^{2k}}$. With the bound $L^{k \NG}$ on the number of possible circuit topologies and $|\Q|^\NG$ combinations for the gates in $\hU_\tveps$, theorem~\ref{theorem:Circuit} follows.

\section{Evolving with time-dependent interactions}\label{sec:tevol}
Similarly, we can bound the volume of operators that is reachable by evolving $\hO$ with respect to time-dependent Hamiltonians $\hH(t)$.
\begin{theorem}\label{theorem:tevol}
  For time-dependent Hamiltonians $\hH(t)=\sum_{i=1}^K\hh_i(t)$ with $K$ terms, let interactions be $k$-local and norm-bounded, i.e., terms $\hh_i(t)$ act on at most $k$ sites and $|h|:=\max_i\sup_{0\leq t\leq T}\|\hh_i(t)\|/\hbar$ is finite. For every term $\hh_i$ and all times $t,s$, let commutators $[\hh_i(t),\hh_j(s)]$ be nonzero for at most $z$ terms $\hh_j$. Observables $\{\hU^\dag(T)\hO\hU(T)\}$, evolved with such Hamiltonians $\{\hH\}$ from $t=0$ to $T$, are elements of
  \begin{subequations}\label{eq:NT}
  \begin{align}
  	\N_T&\leq L^{kK}\left(\frac{112 T^2K^2z|h|^2}{\veps^2}\right)^{4d^{2k} T^2K^2z|h|^2/\veps}\\ \label{eq:NTb}
  	&=L^{kK} e^{\mc{O}\left(T^2K^2z\ln(T^2K^2z)\right)}
  \end{align}
  \end{subequations}
  balls of radius $\veps w(\hO)$ in operator space.
\end{theorem}
The number $K$ of terms in the Hamiltonian is bounded by $\binom{L}{k}\leq L^k$ and we have assumed that the interaction graph (choice of $k$-site supports of interactions terms) is time-independent. The number of interaction graphs is hence bounded by $L^{kK}$. Also, $z$ may be $\mc{O}(L^0)$ but can always be bounded by $kL^{k-1}$ such that $\log \N_T$ is in any case polynomial in the system size $L$ and time $T$.
The decisive step for proving theorem~\ref{theorem:tevol} is a Trotter-Suzuki decomposition \cite{Trotter1959,Suzuki1976,Huyghebaert1990-23} of the time-evolution operator $\hU(t)$ which obeys $i\hbar\partial_t\hU(t)=\hH(t)\hU(t)$ and $\hU(0)=\hid$.
\begin{lemma}\label{lemma:TrotterSuzuki}
  With the preconditions of theorem~\ref{theorem:tevol}, the time-evolution operator can be approximated by the decomposition
  $\hU^{\Delta t}(T):=\prod_{n=1}^{N_t}\prod_{i}\hu_i(n)$
  into $N_t$ time steps of size $\Delta t=T/N_t$, where $\hu_i(n)$ denotes the time-evolution operator from time $(n-1)\Delta t$ to $n\Delta t$, generated by $\hh_i(t)$. The accuracy is
  \begin{equation}\label{eq:TS-error}
  	\|\hU^{\Delta t}(T)-\hU(T)\|\leq \Delta t \,T K z |h|^2.
  \end{equation}
\end{lemma}
\begin{figure}[t]
  \includegraphics[width=\columnwidth]{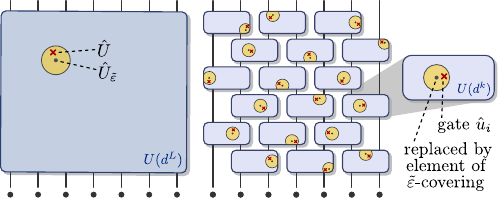}
  \caption{
  Using a Trotter-Suzuki decomposition, the evolution with respect to $k$-local Hamiltonians with arbitrary time-dependence can be approximated by quantum circuits. Each $k$-site gate (here, $k=2$) in a quantum circuit can be approximated by an element of an $\tveps$-covering for the unitary group $U(d^k)$. This construction allows us to bound covering numbers for sets of evolved observables $\{\hU^\dag\hO\hU\}$.}
\end{figure}

This can be shown following the derivation in Ref.~\cite{Huyghebaert1990-23} and applying the triangle inequality. Accuracy $\|\hU^{\Delta t}(T)-\hU(T)\|\leq\veps/4$ is achieved for $N_t=4T^2Kz|h|^2/\veps$ time steps. Now we are basically back to the case of observables that are evolved by a quantum circuit and can proceed, as before, by approximating each of the $KN_t$ gates in $\hU^{\Delta t}$ by the nearest element of an $\tveps$-covering $\mc{Q}$ for $U(d^k)$. Calling the resulting circuit $\hU^{\Delta t}_{\tveps}$, we can achieve accuracy $\|\hU^{\Delta t}_{\tveps}(T)-\hU^{\Delta t}(T)\|\leq\veps/4$ with $|\mc{Q}|\leq(28KN_t/\veps)^{d^{2k}}$ according to lemma~\ref{lemma:CoverUn}. Then,
\begin{equation}
	\|[\hU^{\Delta t}_\tveps(T)]^\dag\hO\hU^{\Delta t}_\tveps(T)-\hU^\dag(T)\hO\hU(T)\|\leq \veps w(\hO)
\end{equation}
and with $|\mc{Q}|^{KN_t}$ combinations for the gates in $\hU^{\Delta t}_\tveps$, theorem~\ref{theorem:tevol} follows.

\section{Packings for projection operators}\label{sec:projections}
To quantify the efficiency with which dynamics explore the set of observables, let us first focus on the case where the accessible observable $\hO$ is a projection operator. Examples for such observables are Pauli measurements that are the standard choice in quantum computing and particle densities that are typical for ultracold-atom experiments.

Let $G_{n,m}$ denote the set of all rank-$n$ projection operators on an $m$-dimensional Hilbert space $\H$, where in our case $m=d^L$. We will determine bounds on covering numbers for $G_{n,m}$ and compare them to Eqs.~\eqref{eq:Ncircuit} and \eqref{eq:NT}.
$G_{n,m}$ can be identified with the Grassmann manifold, the space of all $n$-dimensional subspaces of $\H$, where each such subspace corresponds to the projection onto that subspace. More useful for our purposes, $G_{n,m}$ can also be identified with the quotient group
\begin{equation}\label{eq:G-as-quotient}
	G_{n,m} \cong U(m)/U(n,m),
\end{equation}
where $U(n,m):=U(n)\times U(n-m)$ is the direct product of the unitary groups $U(n)$ and $U(n-m)$. Eq.~\eqref{eq:G-as-quotient} is due to the fact that every $n$-dimensional subspace $\Ht$ of $\H$ can be specified by a fixed reference subspace $\H_0$ of dimension $n$ and an element $\hV$ of $U(m)/U(n,m)$ such that 
\begin{equation}
	\hV(\H_0\oplus\H_0^\bot)=\Ht\oplus\Ht^\bot,
\end{equation}
where $\H_0^\bot$ and $\Ht^\bot$ are the orthogonal complements of $\H_0$ and $\Ht$ in $\H$. Clearly, $\Ht\oplus\Ht^\bot$ is invariant under transformations from $U(n,m)$, which explains the identification \eqref{eq:G-as-quotient}.

Bounds on covering numbers for $U(m)/U(n,m)$ can be obtained from covering numbers of $U(m)$ and $U(n,m)$. Those of $U(n,m)$ can be obtained from covering numbers of $U(m)$ and $U(n)$. In general, we have the following.
\begin{lemma}\label{lemma:CoverProduct}
  Let $(\M_1,d_1)$ and $(\M_2,d_2)$ be metric spaces and $(\M,d):=(\M_1\times\M_2,d_1\times d_2)$ be their direct product with $d\big((x_1,x_2),(y_1,y_2)\big)\equiv\max\{d_1(x_1,y_1),d_2(x_2,y_2)\}$. Then, their covering numbers obey
  \begin{multline}
  	\N(\M_1,d_1,2\veps)\N(\M_2,d_2,2\veps)
  	\leq \N(\M,d,\veps)\\
  	\leq \N(\M_1,d_1,\veps)\N(\M_2,d_2,\veps).
  \end{multline}
\end{lemma}
\begin{lemma}\label{lemma:CoverQuotient}
  Let $G$ be a group and $H$ a compact subgroup, $d$ an invariant metric on $G$ and $d'$ the induced quotient metric on $G/H$,
  \begin{equation}\label{eq:quotientMetric}
  	d'([x],[y])\equiv\inf\{d(e,z)\,|\,z\in G:[y]=[z\cdot x]\}.
  \end{equation}
  Here $e$ denotes the neutral element in $G$ and $[x]\equiv x\cdot H$ the coset of $x$. Then the covering number of $(G/H,d')$ obeys
  \begin{multline}
  	\frac{\N(G,d,2\veps)}{\N(H,d,\veps)}
  	\leq \N(G/H,d',\veps)
  	\leq \frac{\N(G,d,\veps/2)}{\N(H,d,\veps)}.
  \end{multline}
\end{lemma}
In combination with lemma~\ref{lemma:CoverUn} this gives
\begin{equation*}
	\frac{1}{19^{m^2}}\left(\frac{7}{\veps}\right)^{2n(m-n)}\!\!\!\!\!
	\leq \N(G_{n,m},d',\veps)
	\leq 38^{m^2}\!\left(\frac{3}{8\veps}\right)^{2n(m-n)}
\end{equation*}
for covering numbers of the Grassmannians \eqref{eq:G-as-quotient} with $\veps\leq 1/20$. In this case, the induced quotient metric \eqref{eq:quotientMetric} is
	$d'(\H_1,\H_2)=\inf\{\|\hid-\hV\|\,|\, \hV\in U(m)\ \text{with}\ \H_2=\hV\H_1\}$
for all $\H_1,\H_2\in G_{n,m}$ \cite{Szarek1981}. However, we are actually interested in $G_{n,m}$, interpreted as the set of all rank-$n$ projection operators on $\H$. Then the relevant metric is not $d'$ but the operator norm distance $\|\hP_1-\hP_2\|$. So, in the final step, we relate covering numbers for $(G_{n,m},\|\cdot\|)$ to those of $(G_{n,m},d')$.
\begin{lemma}\label{lemma:CoverLipschitz}
  Let $(\M_1,d_1)$ and $(\M_2,d_2)$ be metric spaces and $f:\M_1\to M_2$ bi-Lipschitz such that $f(\M_1)=\M_2$ with
  \begin{alignat*}{4}
  	d_2(f(x),f(y))&\leq K d_1(x,y)\,&&\,\forall x,y\in\M_1\,\,\text{and}\\
  	d_2(f(x),f(y))&\geq k\, d_1(x,y)  &&\forall x,y\in\M_1\,\text{with}\,\,\,d_1(x,y)\leq r.
  \end{alignat*}
  Then, their covering numbers obey
  \begin{equation*}
  	\N(\M_1,d_1,{2\veps}/{k})
  	\leq \N(\M_2,d_2,\veps)
  	\leq \N(\M_1,d_1,{\veps}/{K}),
  \end{equation*}
  where the left inequality requires $\veps\leq kr/2$.
\end{lemma}
As shown in Appx.~\ref{sec:QuotientMetric}, $\sqrt{2}\,d'(\H_1,\H_2)/5\leq\|\hP_1-\hP_2\|\leq 2\,d'(\H_1,\H_2)$ for subspaces $\H_1,\H_2\in G_{n,m}$ which are identified with the projections $\hP_1$ and $\hP_2$ onto these subspaces. Hence, we can apply lemma~\ref{lemma:CoverLipschitz} to $G_{n,m}$ with $d_1$ and $d_2$ being the quotient metric and operator-norm distance, respectively, $K=2$, $k=\sqrt{2}/5$, and $r=2/5$.
\begin{theorem}\label{theorem:CoverGrassmann}
  The $\veps$-covering numbers $\N_G$ for rank-$n$ projection operators on an $m$-dimensional Hilbert space with respect to the operator-norm distance $\|\cdot\|$ obey
  \begin{equation}
  	\frac{1}{19^{m^2}}\left(\frac{9}{5\veps}\right)^{2n(m-n)}\!\!
  	\leq \N_G 
  	\leq 38^{m^2}\!\left(\frac{3}{4\veps}\right)^{2n(m-n)},
  \end{equation}
  where the lower bound is valid for $\veps\leq 1/71$ and the upper one for $\veps\leq 1/10$.
\end{theorem}
The Hilbert space dimension of our many-body systems grows exponentially in the system size, $m=\dim\H=d^L$. For the case of interest, where $n$ and $m-n$ are finite fractions of $m$, i.e., projection operators as those of Pauli measurements ($n=D/2$), and sufficiently small $\veps=\mc{O}(1)$, theorem~\ref{theorem:CoverGrassmann} states that covering numbers for $G_{n,m}$ grow superexponentially with $L$, $\N_G=\exp\left[\Omega(m^2)\right]=\exp\left[\Omega(d^{2L})\right]$. In combination with theorems~\ref{theorem:Circuit} and \ref{theorem:tevol}, this shows that even with full control over the quantum system, generic projections can only be realized by implementing exponential-depth quantum circuits or evolving the system for a time $T$ that grows exponentially with the system size $L$.

\section{Packings for generic observables}\label{sec:observables}
\begin{figure}[t]
  \includegraphics[width=0.97\columnwidth]{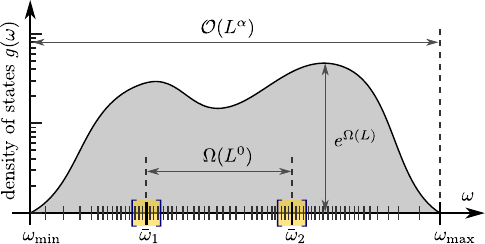}
  \caption{The results on limitations for measurements of projection operators are generalized in Theorem~\ref{theorem:GeneralObs}. It covers observables with a spectral width that grows polynomially with increasing system size $L$, i.e., $w=\mc{O}(L^\alpha)$. There necessarily exist points with an exponential density of states $g(\omega)=e^{\Omega(L)}$. Theorem~\ref{theorem:GeneralObs} applies to observables which have two such points $\bomega_{1,2}$ with distance $\bomega_2-\bomega_1=\Omega(L^0)$.}
\end{figure}
So far, we have only considered observables being projection operators ($w(\hO)=1/2$) and found that measuring them is in general inefficient with respect to growing system size $L$. We can easily extend this result to observables $\hO$ that have only two eigenvalues $\omega_1<\omega_2$ with exponential degeneracies $n$ and $m-n=e^{\Omega(L)}$. As long as the spectral width is polynomial in $L$, $w(\hO)=\frac{\omega_2-\omega_1}{2}=\mc{O}(L^\alpha)$ for some constant $\alpha\geq 0$, theorems~\ref{theorem:Circuit} and \ref{theorem:tevol} with polynomial $N_g$ and $T$ still yield exponential upper bounds on $\veps$-covering numbers for observables that can be reached through evolution of a predefined reference observable. And, as long as $\omega_2-\omega_1$ has an $L$-independent lower bound [is $\Omega(L^0)$], theorem~\ref{theorem:CoverGrassmann} with sufficiently small $\veps$ ($\veps\to\veps/|\omega_2-\omega_1|$) yields superexponential lower bounds on $\veps$-covering numbers for the set of observables with the given spectrum.
In fact, we can generalize much further
\begin{theorem}\label{theorem:GeneralObs}
  For a fixed $\alpha>0$, sufficiently small $\veps>0$, and every system size $L$, let $G_\omega$ be the set of observables with some spectrum $\{\omega_k\}$ of polynomial width $w=\mc{O}(L^\alpha)$. For some $\bomega_1<\bomega_2$ with $\bomega_2-\bomega_1=\Omega(L^0)$, let the $\veps/2$-neighborhoods of $\bomega_1$ and $\bomega_2$ contain exponentially many eigenvalues $\omega_k$, i.e., $\big|\{\omega_k\ \text{with}\ |\omega_k-\bomega_i|\leq\veps/2 \}\big|=e^{\Omega(L)}$. Then $\veps$-covering numbers for $G_\omega$ grow superexponentially in $L$ and, generally, elements of $G_\omega$ cannot be reached through application of polynomial-depth quantum circuits or polynomial-time evolution with Hamiltonians as characterized in theorems~\ref{theorem:Circuit} and \ref{theorem:tevol}.
\end{theorem}
This is because, for every observable $\hO\in G_\omega$, we can define $\hO'$ by replacing all eigenvalues in the $\veps/2$-neighborhood of $\bomega_i$ by $\bomega_i$. So, eigenvalues $\bomega_1$ and $\bomega_2$ of $\hO'$ have exponential degeneracies $n,m-n=e^{\Omega(L)}$. For sufficiently small $\veps$, theorem~\ref{theorem:CoverGrassmann} now yields superexponential lower bounds on $\veps/2$-covering numbers for the set of operators that differ from $\hO'$ only in terms of the $\bomega_i$-eigenvectors. As $\|\hO-\hO'\|\leq \veps/2$, it follows that $\veps$-covering numbers for $G_\omega$ also grow superexponentially in $L$.

\section{Discussion}
Theorem~\ref{theorem:GeneralObs} accounts for all typical classes of observables: (a) projection operators as, for example, occurring in Pauli measurements, (b) observables that act in a finite-size subspace like single-site observables $\hO_i$ or two-site operators $\hO_i\hO_j$ for two-point correlation functions, (c) extensive observables like energy etc. As a matter of fact, observables $\hO$ with a polynomial spectral width $w(\hO)$, usually obey the preconditions of theorem~\ref{theorem:GeneralObs}: Due to the exponential growth of the Hilbert space with $L$, the density of eigenstates for such observables generally grows exponentially in the bulk of the spectrum and, hence, points $\bomega_1$ and $\bomega_2$ with the required properties generally exist.

Hence, dynamical measurement schemes for observables that are not directly accessible, i.e., a controlled time evolution and subsequent measurement of directly accessible observables, are in general inefficient. For a predefined accuracy, generally, the required evolution time increases exponentially with the system size. So it is a question of clever design to allow for the measurement of observables of interest through efficient dynamical schemes.

\acknowledgments
TB thanks Juri Barthel, Kenneth Brown, Dripto Debroy, and Jungsang Kim for helpful discussions. JL is supported
in part by the National Science Foundation under grant DMS-1454939.

\appendix

\section{Proof of lemma~\ref{lemma:TrotterSuzuki} -- Trotter-Suzuki}\label{sec:TrotterSuzuki}
Let $\hU^{t,s}_H$ denote the time-evolution operator for the evolution from time $s$ to time $t\geq s$ under a time-dependent Hamiltonian $\hH(t)$ and let $\hU^{s,t}_H:=(\hU^{t,s}_H)^\dag$. The time-evolution operator obeys the equations of motion
\begin{equation*}
	i\hbar\partial_t\hU^{t,s}_H=\hH(t)\hU^{t,s}_H,\quad
	i\hbar\partial_s\hU^{t,s}_H=-\hU^{t,s}_H \hH(s),
\end{equation*}
and $\hU^{t,t}_H=\hid$ $\forall$ $t$.

Let us first recapitulate a result by Huyghebaert and De Raedt \cite{Huyghebaert1990-23} that bounds the error for approximating $\hU^{t,s}_{H+h}$, i.e., the evolution under two Hamiltonian terms $\hH(t)+\hh(t)$, by the product $\hU^{t,s}_H\hU^{t,s}_h$. The operator-norm distance is
\begin{align}\nonumber
	\big\|\hU^{t,q}_{H+h}-\hU^{t,q}_H&\hU^{t,q}_h\big\|
	=\big\|\hU^{t,q}_{H+h}\hU^{q,t}_h\hU^{q,t}_H-\hid\big\|\\\nonumber
	&\textstyle\leq \int_q^t\ud r\,\big\|\partial_r(-\hU^{t,r}_{H+h}\hU^{r,t}_h\hU^{r,t}_H)\big\|\\\nonumber
	&\textstyle= \frac{1}{\hbar}\int_q^t\ud r\,\big\|\hH(r)-\hU^{r,t}_h\hH(r)\hU^{t,r}_h\big\|\\\nonumber
	&\textstyle\leq \frac{1}{\hbar} \int_q^t\ud r\,\int_r^t\ud s\,\big\|\partial_s(-\hU^{r,s}_h\hH(r)\hU^{s,r}_h)\big\|\\\label{eq:TS-step1Term}
	&\textstyle\leq \frac{1}{\hbar^2} \int_q^t\ud r\,\int_r^t\ud s\,\big\|[\hh(s),\hH(r)]\big\|.
\end{align}
Here, we have employed the invariance of the norm under unitary transformations, the fundamental theorem of calculus, and the triangle inequality.

Theorem~\ref{theorem:tevol} and lemma~\ref{lemma:TrotterSuzuki} address Hamiltonians $\hH(t)=\sum_{i=1}^K\hh_i(t)$ with $K$ norm-bounded terms $\hh_i(t)$. Let $|h|:=\frac{1}{\hbar}\max_i\sup_{q\leq s\leq t}\|\hh_i(s)\|$, and for every term $\hh_i$ and all times $s,r\in[q,t]$, let commutators $[\hh_i(s),\hh_j(r)]$ be nonzero for at most $z$ terms $\hh_j$. Then, $\frac{1}{\hbar^2}\big\|[\hh_i(s),\hH(r)]\big\|\leq 2z|h|^2$. Applying Eq.~\eqref{eq:TS-step1Term} and the triangle inequality iteratively, we obtain
\begin{equation}\label{eq:TS-step}\textstyle
	\big\|\hU^{t,q}_{H}-\prod_{i=1}^K\hU^{t,q}_{h_i}\big\|
	\leq (t-q)^2Kz|h|^2.
\end{equation}

Finally, for an evolution from time $t=0$ to time $T$, the time window can be split into $N_t$ steps of size $\Delta t=T/N_t$. Applying the triangle inequality and Eq.~\eqref{eq:TS-step} for each time step $\hU^{n\Delta t,(n-1)\Delta t}_{H}$, one arrives at the error bound Eq.~\eqref{eq:TS-error}, stated in lemma~\ref{lemma:TrotterSuzuki} for the the Trotter-Suzuki approximation
\begin{equation}
	\hU_H^{\Delta t}(T)=\prod_{n=1}^{N_t}\prod_{i=1}^K\hU^{n\Delta t,(n-1)\Delta t}_{h_i}
	\,\approx\, \hU^{T,0}_H.
\end{equation}

\section{Covering and packing numbers}\label{sec:CoveringsPackings}
Let $(\M,d)$ be a metric space.
A subset $\Q\subset\M$ is called an \emph{$\veps$-covering} if $\forall$ $z\in\M$ $\exists$ $x\in\Q$ with $d(x,z)\leq\veps$. Then the union of $\veps$-balls around the elements of $\Q$ covers $\M$. The cardinality $\N(\M,d,\veps)$ of the smallest $\veps$-covering is called the \emph{$\veps$-covering number} of $(\M,d)$.

A subset $\Q\subset\M$ is called an \emph{$\veps$-packing} if $d(x,y)>\veps$ $\forall$ $x\neq y\in\Q$. Then $\veps/2$-balls around the elements of $\Q$ are all disjoint. The cardinality $\bN(\M,d,\veps)$ of the largest $\veps$-packing is called the \emph{$\veps$-packing number} of $(\M,d)$.

Covering and packing numbers obey the well-known inequalities \cite{Kolmogorov1959-14}
\begin{equation}\label{eq:CoverVsPack}
	\bN(\M,d,2\veps)\leq \N(\M,d,\veps)\leq\bN(\M,d,\veps).
\end{equation}
Suppose there exists an $\veps$-covering $\Q$ with cardinality $N$ and a $2\veps$-packing $\Q'$ with cardinality $\bar{N}>N$ for $(\M,d)$. Then, there must exist (at least) two elements $y_1,y_2\in\Q'$ that are both contained in the $\veps$-ball around some $x\in\Q$. Consequently, their distance $d(y_1,y_2)$ cannot be larger than $2\veps$. This contradiction proves the left inequality in Eq.~\eqref{eq:CoverVsPack}.

If $\Q$ is a maximal $\veps$-packing for $(\M,d)$, we cannot find a point in $\M$ with distance larger than $\veps$ from all points in $\Q$. Hence, $\Q$ is also an $\veps$-covering, which proves the right inequality in Eq.~\eqref{eq:CoverVsPack}. This also shows the existence of \emph{$\veps$-nets}, which are simultaneously $\veps$-coverings and $\veps$-packings.

\section{Proof of lemma~\ref{lemma:CoverLipschitz} -- Two metric spaces}\label{sec:TwoMetricSpaces}
Lemma~\ref{lemma:CoverLipschitz} uses covering numbers for a metric space $(\M_1,d_1)$ to bound covering numbers for a second metric space $(\M_2,d_2)$. The precondition is that there exists a bi-Lipschitz function $f$ with $f(\M_1)=\M_2$ and 
\begin{alignat*}{4}
	d_2(f(x),f(y))&\leq K d_1(x,y)\,&&\,\forall x,y\in\M_1\,\,\text{and}\\
	d_2(f(x),f(y))&\geq k\, d_1(x,y)  &&\forall x,y\in\M_1\,\text{with}\,\,d_1(x,y)\leq r.
\end{alignat*}

Let $\Q$ be an $\veps/K$-covering for $(\M_1,d_1)$. This implies that $\forall$ $z\in\M_1$ $\exists$ $x\in\Q$ such that $d_1(x,z)\leq\veps/K$. It follows that $\forall$ $z'\in\M_2$ $\exists$ $x'\in f(\Q)$ such that $d_2(x',z')\leq K d_1(x,z)\leq \veps$, where $x$ and $z$ have been chosen such that $f(x)=x'$ and $f(z)=z'$. So, $f(\Q)$ is an $\veps$-covering of $(\M_2,d_2)$ and
\begin{equation}
	\N(\M_2,d_2,\veps) \leq \N(\M_1,d_1,{\veps}/{K}).
\end{equation}

Now, let $\Q$ be a $2\veps/k$-packing for $(\M_1,d_1)$. This implies that $d_1(x,y)> 2\veps/k$ $\forall$ $x\neq y\in\Q$. It follows that $\forall$ $x'\neq y'\in f(\Q)$, we have $d_2(x',y')\geq k\, d_1(x,y)>2\veps$. Here, $x$ and $y$ have been chosen such that $f(x)=x'$ and $f(y)=y'$. Further, we have assumed that $d_1(x,y)\leq r$ for the relevant $x$ and $y$, which corresponds to the upper bound $kr/2$ on the allowed $\veps$.
So, $f(\Q)$ is a $2\veps$-packing for $(\M_2,d_2)$ and the packing numbers obey $\bN(\M_2,d_2,2\veps)\geq\bN(\M_1,d_1,2\veps/k)$. Application of Eq.~\eqref{eq:CoverVsPack} then gives
\begin{equation}
	\N(\M_2,d_2,\veps) \geq \N(\M_1,d_1,{2\veps}/{k})
\end{equation}
for the covering numbers.

\section{Proof of lemma~\ref{lemma:CoverUn} -- Covering \texorpdfstring{$U(n)$}{U(n)}}\label{sec:Un}
Lemma~\ref{lemma:CoverUn} bounds covering numbers for $U(n)$, the group of unitary $n\times n$ matrices. It can be proven in two steps. Any unitary can be obtained by exponentiating an element of its Lie algebra $u(n)$. Actually, a ball of radius $\pi$ in $u(n)$ is sufficient. Then, bounds on covering numbers for this ball in $u(n)$ and Lipschitz constants for the exponential map in conjunction with lemma~\ref{lemma:CoverLipschitz} allow us to prove lemma~\ref{lemma:CoverUn}.

First, note that $\veps$-covering numbers for balls of radius $R$ in $D$ dimensions, $\B_R\equiv R\cdot\B_1\subset\mathbb{R}^D$ obey
\begin{equation}\label{eq:CoverBall}
	\left(\frac{R}{\veps}\right)^D
	\leq \N(\B_R,\|\cdot\|,\veps)
	\leq \left(1+\frac{2R}{\veps}\right)^D
\end{equation}
This standard result follows from a comparison of volumes.
Let $\Q\subset\B_R$ be an $\veps$-packing for $\B_R$. As balls of radius $\veps/2$ around the points in $\Q$ are disjoint and fully contained in the ball $\B_{R+\veps/2}$, the upper bound in Eq.~\eqref{eq:CoverBall} follows by comparing the corresponding volumes,
\begin{equation*}
	|\Q|\leq\frac{\vol \B_{R+\veps/2}}{\vol \B_{\veps/2}}
	=\frac{(R+\veps/2)^D\vol\B_1}{(\veps/2)^D\vol\B_1}
	=\left(1+\frac{2R}{\veps}\right)^D\!.
\end{equation*}
For the lower bound in Eq.~\eqref{eq:CoverBall}, consider an $\veps$-covering $\Q$ of $\B_R$. Then $\B_R$ is fully contained in the union of $\veps$-balls around points in $\Q$, $\B_R\subset\Q+\B_\veps$. Hence,
\begin{equation*}
	|\Q|\geq\frac{\vol \B_R}{\vol \B_\veps}=\left(\frac{R}{\veps}\right)^D.
\end{equation*}

Secondly, note that $U(n)$ is obtained by exponentiation of skew-Hermitian matrices $J$ with operator norm $\|J\|\leq \pi$, i.e.,
\begin{equation}\label{eq:Un_is_exp_un}
	U(n)=\exp[\B_\pi(u(n))].
\end{equation}
This is so, because unitary matrices $U\in U(n)$ are normal  with eigenvalues $e^{i\lambda_k}$ of amplitude 1. Hence, $U=V^\dag \diag(e^{i\lambda_1},\dotsc,e^{i\lambda_n})V$ with unitary $V$ and $U=\exp(J)$ with $J=V^\dag \diag(i\lambda_1,\dotsc,i\lambda_n)V\in u(n)$. Now, as we can choose $|\lambda_k|\leq\pi$ such that $\|J\|\leq\pi$, Eq.~\eqref{eq:Un_is_exp_un} follows.

We want to apply lemma~\ref{lemma:CoverLipschitz} to obtain covering numbers for $U(n)$ based on covering numbers for $\B_\pi(u(n))$. To this purpose, we need Lipschitz constants for the exponentiation of skew-Hermitian matrices. We claim that for $X,Y\in u(n)$,
\begin{equation}\label{eq:UnLipschitz}
	(2-e^r)\|X-Y\|\leq\big\|e^X-e^Y\big\|\leq \|X-Y\|,
\end{equation}
where the left inequality is valid for $\|X\|,\|Y\|\leq r$.
The upper bound in Eq.~\eqref{eq:UnLipschitz} follows by writing $e^X-e^Y$ as a telescope sum
\begin{equation*}
	e^X-e^Y = \sum_{k=1}^m e^{(k-1)\frac{X}{m}}\left(e^{\frac{X}{m}}-e^{\frac{Y}{m}}\right)e^{(m-k)\frac{Y}{m}}\quad\forall m
\end{equation*}
and using the triangle inequality and the invariance of the operator norm under unitary transformations like $e^{\alpha X}$ and $e^{\alpha Y}$,
\begin{equation*}
	\big\|e^X-e^Y\big\|\leq \lim_{m\to\infty} m\big\|e^{X/m}-e^{Y/m}\big\|=\|X-Y\|.
\end{equation*}
Assuming $\|X\|,\|Y\|\leq r$, the lower bound in Eq.~\eqref{eq:UnLipschitz} can be derived by Taylor expansion
\begin{equation*}
	\big\|e^X-e^Y\big\|\geq \|X-Y\|-\Big\|\sum_{k\geq 2}\frac{1}{k!}(X^k-Y^k)\Big\|
\end{equation*}
and using that
\begin{align*}
	\Big\|\sum_{k\geq 2}&\frac{1}{k!}(X^k-Y^k)\Big\|
	=\Big\|\sum_{k\geq 2}\sum_{\ell=1}^k\frac{1}{k!} X^{\ell-1}(X-Y)Y^{k-\ell}\Big\|\\
	&\leq\sum_{k\geq 2}\frac{1}{(k-1)!}\,r^{k-1}\|X-Y\|
	=(e^r-1)\|X-Y\|.
\end{align*}

With Eqs.~\eqref{eq:CoverBall}--\eqref{eq:UnLipschitz} the preconditions of lemma~\ref{lemma:CoverLipschitz} are fulfilled for $\M_1=\B_\pi(u(n))$ and $\M_2=U(n)$, $d_1$ and $d_2$ being the operator-norm distance, and $f(X)=e^X$. The upper bound in Eq.~\eqref{eq:UnLipschitz} gives $K=1$ and, choosing $r=2/5$, the lower bound gives $2-e^r>1/2=:k$. As $u(n)$ corresponds to an $n^2$-dimensional real vector space, $D=n^2$ and $R=\pi$ when we apply Eq.~\eqref{eq:CoverBall} for $\B_\pi(u(n))$. Lemma~\ref{lemma:CoverLipschitz} then yields
\begin{multline}
	 \left(\frac{\pi}{4\veps}\right)^{n^2}\!
	 =\left(\frac{kR}{2\veps}\right)^D\!
	 \leq \N\left(U(n),\|\cdot\|,\tveps\right)\\
	 \leq \left(1+\frac{2KR}{\veps}\right)^D\!
	 = \left(1+\frac{2\pi}{\veps}\right)^{n^2}.
\end{multline}
The left inequality requires $\veps\leq kr/2=1/10$. Assuming the same constraint for the right-hand side, we have $1+{2\pi}/{\veps}\leq 7/\veps$ and, thus, lemma~\ref{lemma:CoverUn}.

\section{Proof of lemma~\ref{lemma:CoverProduct} -- Covering direct products}\label{sec:Product}
Lemma~\ref{lemma:CoverProduct} uses covering numbers for two metric spaces $(\M_1,d_1)$ and $(\M_1,d_2)$ to bound covering numbers for their direct product $(\M,d):=(\M_1\times\M_2,d_1\times d_2)$, where $d\big((x_1,x_2),(y_1,y_2)\big)\equiv\max\{d_1(x_1,y_1),d_2(x_2,y_2)\}$. 

Let $\Q_1$ and $\Q_2$ be $\veps$-coverings for $(\M_1,d_1)$ and $(\M_2,d_2)$, respectively. Then $\forall$ $(z_1,z_2)\in\M$ $\exists$ $(x_1,x_2)\in\Q:=\Q_1\times\Q_2$ such that $d\big((x_1,x_2),(z_1,z_2)\big)\leq\veps$. So, $\Q$ is an $\veps$-covering for $(\M,d)$ and
\begin{equation}
	\N(\M,d,\veps) \leq \N(\M_1,d_1,\veps)\,\N(\M_2,d_2,\veps).
\end{equation}

Now, let $\Q_1$ and $\Q_2$ be $2\veps$-packings for $(\M_1,d_1)$ and $(\M_2,d_2)$, respectively. Then $d(x,y)> 2\veps$ $\forall$ $x\neq y\in\Q:=\Q_1\times\Q_2$. So, $\Q$ is a $2\veps$-packing for $(\M,d)$ and, with Eq.~\eqref{eq:CoverVsPack},
\begin{equation}
	\N(\M,d,\veps) \geq \N(\M_1,d_1,2\veps)\,\N(\M_2,d_2,2\veps).
\end{equation}

\section{Proof of lemma~\ref{lemma:CoverQuotient} -- Covering quotient groups}\label{sec:Quotient}
Let $G$ be a group with an invariant metric $d$ and a compact subgroup $H$. We refer to elements of the quotient group $G/H$ by $[x]:=x\cdot H$ $\forall$ $x\in G$. The induced quotient metric [cf. Eq.~\eqref{eq:quotientMetric}]
\begin{equation}\label{eq:quotientMetric2}
  	d'([x],[y])\equiv\inf\{d(x',y')\,|\,x'\in[x],\,y'\in [y]\}
\end{equation}
is a valid metric on $G/H$. Lemma~\ref{lemma:CoverQuotient} uses covering numbers for $(G,d)$ and $(H,d)$ to bound covering numbers for $(G/H,d')$. In the following, we will consider different subsets $\A$ of $G/H$. Choosing for every element of such a subset $\A$ an arbitrary representative $a\in G$ and denoting the set of these representatives $\A_r$, one has $\A=\{[a]\,|\,a\in\A_r\}$ with equal cardinalities $|\A_r|=|\A|$.

Let $\A$ and $\B$ be $\veps$-coverings for $(G/H,d')$ and $(H,d)$, respectively. We will find that $\Q:=\{ab\,|\,a\in\A_r,\,b\in\B\}$ then is a $2\veps$-covering of $(G,d)$.
Note that $\forall$ $z\in G$ $\exists$ $a\in \A_r$ such that $d'([a],[z])\leq \veps$. Choose $h\in H$ such that $d(ah,z)$ is minimized ($\leq\veps$). Now, there exists an element $b\in\B$ with $d(b,h)\leq \veps$ and with the triangle inequality it follows that
\begin{multline*}
 	d(ab,z)=d(b,a^{-1}z)\leq d(b,h)+ d(h,a^{-1}z)\\
 	= d(b,h) + d(ah,z) \leq 2\veps
\end{multline*}
and thus
\begin{equation}
	\N(G,d,2\veps) \leq \N(G/H,d',\veps)\,\N(H,d,\veps).
\end{equation}

Let $\A$ and $\B$ be $\veps$-packings for $(G/H,d')$ and $(H,d)$, respectively. We will find that $\Q:=\{ab\,|\,a\in\A_r,\,b\in\B\}$ then is an $\veps$-packing for $(G,d)$.
Note that $d'([a],[a'])>\veps$ $\forall$ $a\neq a'\in\A_r$. It follows that $d(ah,a'h')>\veps$ $\forall$ $h,h'\in H$ and, hence, $d(ab,a'b')>\veps$ $\forall$ $a,a'\in\A_r$ and $b,b'\in \B$ with either $a\neq a'$ and/or $b\neq b'$.
So, $\Q$ is indeed an $\veps$-packing for $(G,d)$ and, with Eq.~\eqref{eq:CoverVsPack},
\begin{equation}
	\N(G,d,\veps/2) \geq \N(G/H,d',\veps)\,\N(H,d,\veps).
\end{equation}

\section{Quotient metric and operator-norm distance on Grassmannians}\label{sec:QuotientMetric}
The Grassmannian $G_{n,m} \cong U(m)/U(n,m)$ is the space of $n$-dimensional subspaces of an $m$-dimensional Hilbert space. The quotient metric on $G_{n,m}$, induced by the operator-norm distance on the involved unitary groups, is [cf. Eq.~\eqref{eq:quotientMetric}]
\begin{equation}\label{eq:quotientMetricG}
	d'(\H_1,\H_2)=\inf\{\|\hid-\hV\|\,|\, \hV\in U(m)\ \text{with}\ \H_2=\hV\H_1\}
\end{equation}
for all $\H_1,\H_2\in G_{n,m}$. We can identify each element of $G_{n,m}$ with the rank-$n$ projection onto that subspace. In particular, let $\hP$ and $\hQ$ be the projection operators onto some subspaces $\H_1,\H_2\in G_{n,m}$. In the main text, we used lemma~\ref{lemma:CoverLipschitz} to bound covering numbers for $(G_{n,m},\|\cdot\|)$, with $\|\cdot\|$ denoting the operator-norm distance for the projection operators, using covering numbers for $(G_{n,m},d')$. This makes it necessary to derive Lipschitz constants for $\|\hP-\hQ\|$. In the following, we will see that
\begin{equation}\label{eq:quotientMetricGLip}
	\frac{\sqrt{2}}{5}\,d'(\H_1,\H_2)\leq\|\hP-\hQ\|\leq 2\,d'(\H_1,\H_2)
\end{equation}

For the upper bound, let $\hV$ be an optimal unitary in Eq.~\eqref{eq:quotientMetricG}. Then
\begin{multline*}
	\|\hP-\hQ\|=\|\hP-\hV\hP\hV^\dag\|
	=\|\hP-\hV\hP+\hV\hP+\hV\hP\hV^\dag\|\\
	\leq \|(\hid-\hV)\hP\|+\|\hP(\hid-\hV^\dag)\|
	\leq 2\,d'(\H_1,\H_2).
\end{multline*}
as the operator norm is non-increasing under projections.

To derive the lower bound in Eq.~\eqref{eq:quotientMetricGLip}, we can employ a trick developed in the context of perturbation theory \cite{Kato1995}.
Let $\hP$ and $\hQ$ project onto $n$-dimensional subspaces $\H_1$ and $\H_2$, respectively, with $\|\hP-\hQ\|\leq 1/\sqrt{2}$. The operator $\hR:=(\hP-\hQ)^2=\hP+\hQ-\hP\hQ-\hQ\hP$ commutes with $\hP$ and $\hQ$ and
\begin{equation*}
	\hV:=(\hid-\hR)^{-1/2}\hV'
	\,\ \text{with}\,\ 
	\hV'=\hQ\hP+(\hid-\hQ)(\hid-\hP)
\end{equation*}
is a unitary map from $\H_1$ to $\H_2$ \cite{Kato1995}. According to Eq.~\eqref{eq:quotientMetricG}, $\|\hid-\hV\|$ then provides an upper bound on $d'(\H_1,\H_2)$. We can obtain an expression linear in $\|\hP-\hQ\|$ as follows.
\begin{multline}\label{eq:Bound_1-V}\textstyle
	\|\hid-\hV\|\leq \|\hid-\hR\|^{-1/2}\,\big\|\sqrt{\hid-\hR}-\hV'\big\|\\\textstyle
	\leq \sqrt{2}\left(\|\hid-\frac{1}{2}\hR-\hV'\|+\big\|\hid-\frac{1}{2}\hR-\sqrt{\hid-\hR}\big\|\right)
\end{multline}
where we have used that $\|\hid-\hR\|^{1/2}\geq(1-\|\hR\|)^{1/2}\geq 1/\sqrt{2}$ as, by precondition, $\|\hR\|\leq 1/2$. The first term in Eq.~\eqref{eq:Bound_1-V} is $\|\hid-\frac{1}{2}\hR-\hV'\|=\|\hQ(\hQ-\hP)+\frac{1}{2}(\hP-\hQ)\hP+\frac{1}{2}(\hP-\hQ)\hQ\|\leq 2 \|\hP-\hQ\|$. The second term in Eq.~\eqref{eq:Bound_1-V} can be bounded according to $\big\|\hid-\frac{1}{2}\hR-(\hid-\hR)^{1/2}\big\|\leq 1-\frac{1}{2}\|\hR\|-(1-\|\hR\|)^{1/2}\leq \frac{1}{2}\|\hR\|\leq \frac{1}{2}\|\hP-\hQ\|$, such that we arrive at the lower bound for $\|\hP-\hQ\|$ as stated in Eq.~\eqref{eq:quotientMetricGLip}

\bibliographystyle{prsty.tb.title}

\end{document}